\documentclass[a4paper,11pt]{article}
\usepackage{pos}

\newcommand{\be}{\begin{eqnarray}}
\newcommand{\ee}{\end{eqnarray}}

\title{Towards a new generation of reflection models for precision measurements of accreting black holes}
\ShortTitle{Towards a new generation of reflection models}

\author*[a,b]{Cosimo~Bambi}

\affiliation[a]{Center for Astronomy and Astrophysics and Department of Physics, Fudan University,\\
2005 Songhu Road, Shanghai 200438, China}

\affiliation[b]{School of Natural Sciences and Humanities, New Uzbekistan University,\\
Movarounnahr Street 1, Tashkent 100007, Uzbekistan}

\emailAdd{bambi@fudan.edu.cn}

\abstract{Blurred reflection features are commonly observed in the X-ray spectra of accreting black holes. In the presence of high-quality data and with the correct astrophysical model, X-ray reflection spectroscopy is a powerful tool to probe the strong gravity region of black holes, study the morphology of the accreting matter, measure black hole spins, and test Einstein's theory of General Relativity in the strong field regime. In the past 10-15 years, there has been significant progress in the development of the analysis of these reflection features, thanks to both more sophisticated theoretical models and new observational data. However, the next generation of X-ray missions (e.g. \textsl{eXTP}, \textsl{Athena}, \textsl{HEX-P}) promises to provide unprecedented high-quality data, which will necessarily require more accurate synthetic reflection spectra than those available today. In this talk, I will review the state-of-the-art in reflection modeling and I will present current efforts to develop a new generation of reflection models with machine learning techniques.}

\FullConference{Frontier Research in Astrophysics – IV (FRAPWS2024)\\
 9-14 September 2024\\
Mondello, Palermo, Italy\\}

\begin{document}

\maketitle


\section{Introduction}\label{s-1}

\begin{figure}[b]
\centering
\includegraphics[width=0.45\linewidth]{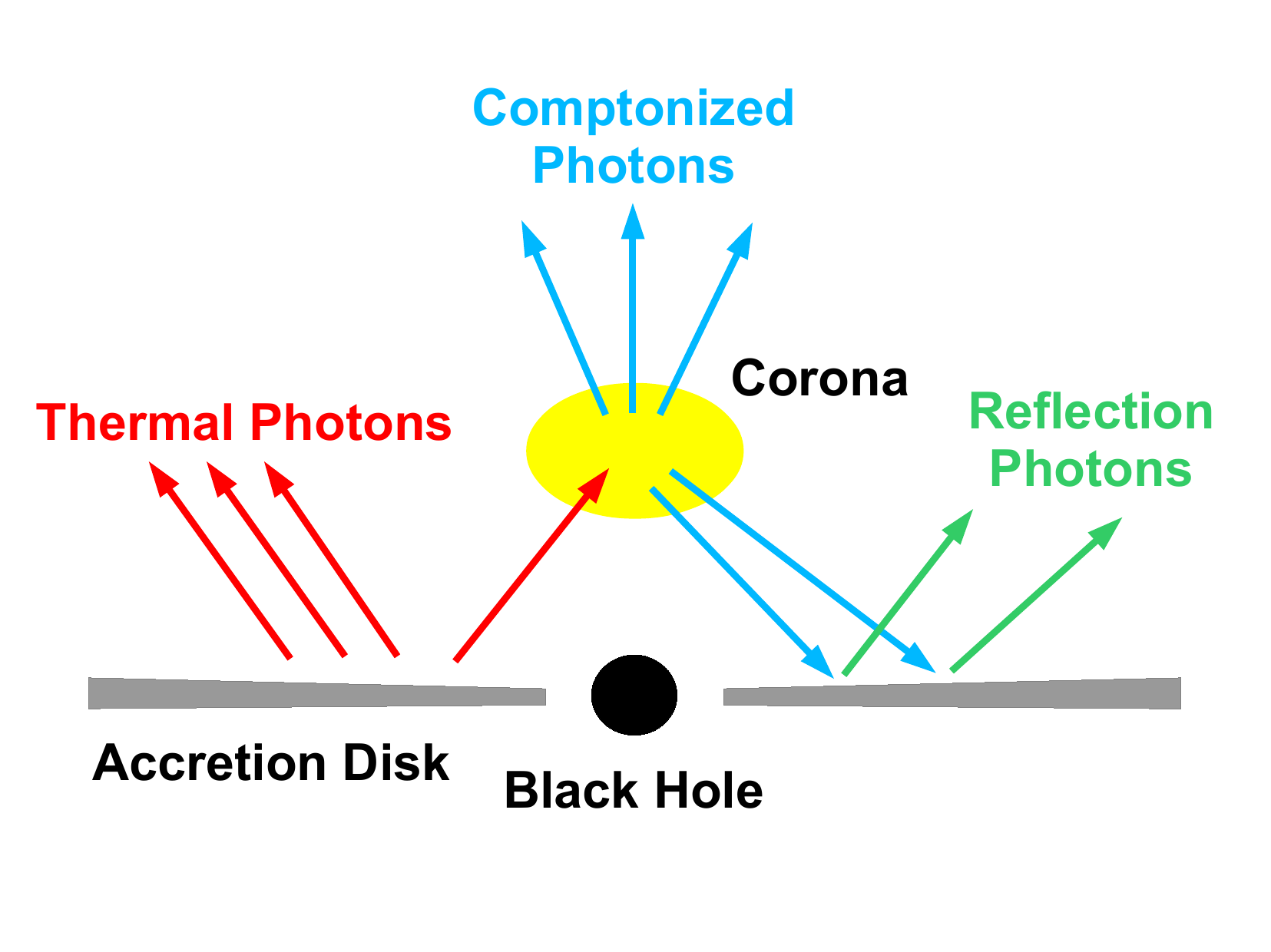}
\hspace{0.5cm}
\includegraphics[width=0.45\linewidth]{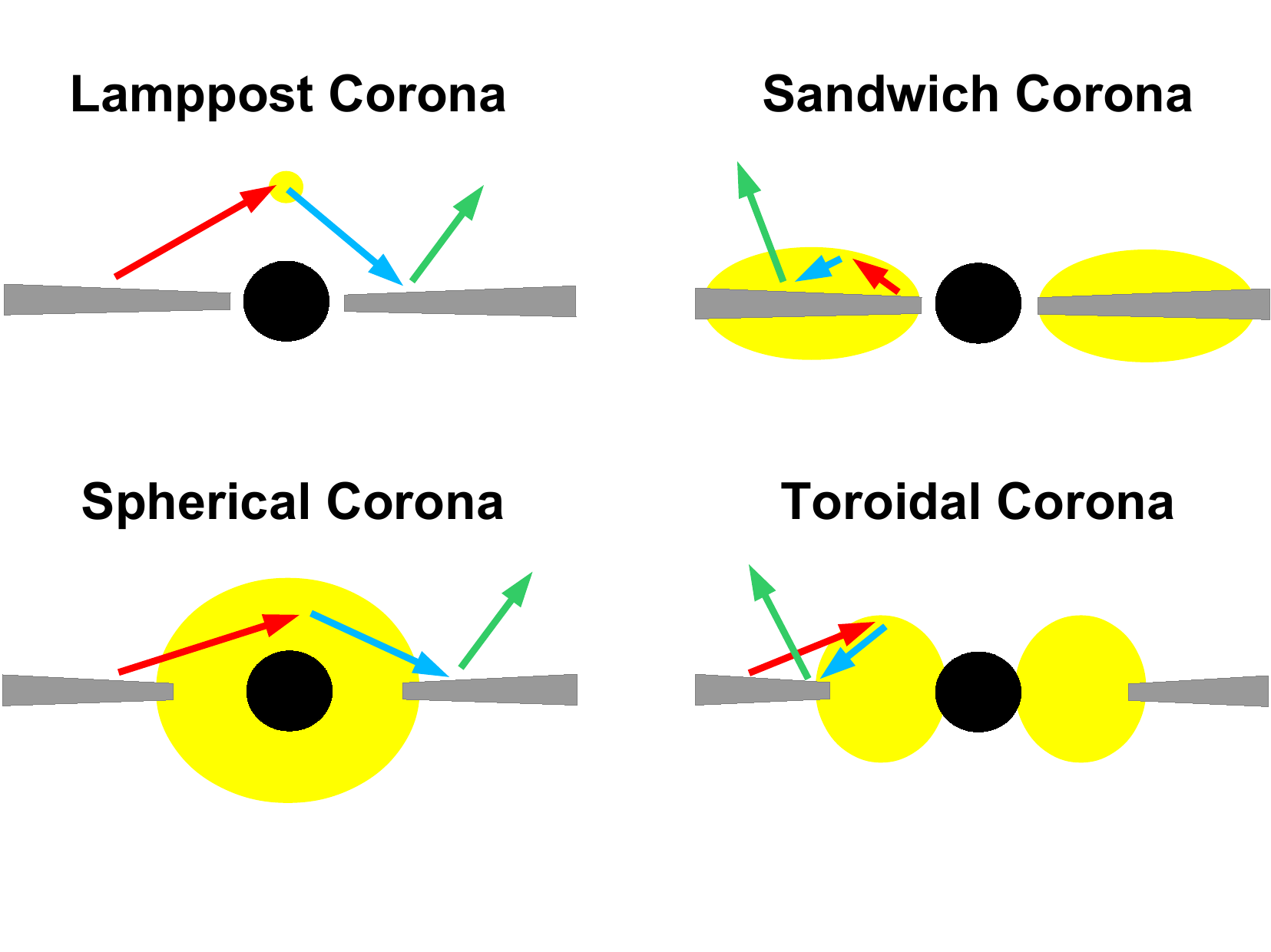}
\vspace{-0.5cm}
\caption{Left panel: Disk-corona model. Right panel: Examples of possible coronal geometries. Figures from Ref.~\cite{Bambi:2024hhi}.}
\label{f-corona}
\end{figure}

Blurred reflection features are commonly observed in the X-ray spectra of stellar-mass black holes in X-ray binaries and supermassive black holes in active galactic nuclei~\cite{Nandra:2007rp}. They are generated by illumination of a cold accretion disk by a hot corona~\cite{Fabian:1989ej,Risaliti:2013cga}. The astrophysical system is sketched in the left panel in Fig.~\ref{f-corona} and is normally referred to as the disk-corona model. The black hole accretes from a cold, geometrically thin, and optically thick accretion disk. The disk is cold because it can efficiently emit radiation. The thermal spectrum of the disk is normally peaked in the soft X-ray band in the case of black hole X-ray binaries and in the UV band in the case of active galactic nuclei~\cite{Shakura:1972te,Page:1974he}. The corona is some hotter material ($\sim 100$~keV) near the black hole and the inner part of the disk, but its exact morphology is not yet understood well~\cite{Galeev:1979td,Haardt:1991tp}. The right panel in Fig.~\ref{f-corona} shows possible coronal geometries. In the lamppost model, the corona is compact and along the black hole spin axis (for instance, the black hole jet may act as a lamppost corona~\cite{Markoff:2005ht}). In the sandwich model, the corona is the atmosphere above the accretion disk. In the spheroidal and toroidal models, the corona is the hot and low-density material in the plunging region, between the inner edge of the disk and the black hole.

Since the disk is ``cold'' and the corona is ``hot'', thermal photons from the disk can inverse Compton scatter off free electrons in the corona. The X-ray spectrum of the Comptonized photons can be normally approximated by a power law with a high-energy cutoff~\cite{Zdziarski:1996wq,Zdziarski:2019cvs}. A fraction of the Comptonized photons can illuminate the disk: Compton scattering and absorption followed by fluorescent emission produce the reflection spectrum.

The {\it non-relativistic reflection spectrum}, namely the reflection spectrum in the rest-frame of the material in the disk, is characterized by narrow fluorescent emission lines in the soft X-ray band and a Compton hump with a peak around 20-40~keV~\cite{Ross:2005dm,Garcia:2010iz}. The strongest emission feature is usually the iron K$\alpha$ line, which is a very narrow feature around 6.4~keV in the case of neutral or weakly ionized iron atoms and shifts up to 6.97~keV in the case of H-like iron ions. The {\it relativistic reflection spectrum}, namely the reflection spectrum of the whole disk detected far from the source, is blurred because of relativistic effects: Doppler boosting because of the motion of the material in the disk and gravitational redshift because the photons have to climb the gravitational potential of the black hole~\cite{Bambi:2017khi}.

X-ray reflection spectroscopy refers to the analysis of these relativistically blurred reflection features. In the presence of high-quality data and the correct astrophysical model, X-ray reflection spectroscopy is a powerful tool to study the strong gravity region of black holes and the morphology of the accreting matter. This technique has been used to measure the spins of about 40~stellar-mass black holes in X-ray binaries and about 40~supermassive black holes in active galactic nuclei, and it is currently the only mature method to measure the spins of supermassive black holes~\cite{Bambi:2020jpe,Draghis:2022ngm,Draghis:2023vzj}. X-ray reflection spectroscopy has also been used to test Einstein's theory of General Relativity in the strong field regime, and currently provides the most stringent tests of the Kerr hypothesis, according to which the spacetime metric around astrophysical black holes is described well by the Kerr solution as predicted by General Relativity~\cite{Bambi:2016sac,Cao:2017kdq,Tripathi:2018lhx,Tripathi:2020yts,Bambi:2015kza}.

The logical way to calculate a reflection spectrum for a specific system can be summarized as follows:
\begin{enumerate}
\item We assume a certain coronal model and we fire photons from the corona to the disk. At the end of these calculations, we have the spectrum of the radiation illuminating every point of the disk. 
\item For every point of the disk, we calculate the reflection spectrum produced by the incident spectrum. Assuming that the system is axisymmetric, at the end of these calculations we have the reflection spectrum at every radial coordinate of the disk.
\item We consider a distant observer and we fire photons from the image plane of the distant observer to the disk. When a photon hits the disk, we calculate the redshift factor $g = E_{\rm o}/E_{\rm e}$, where $E_{\rm o}$ is the photon energy at the detection point on the plane of the distant observer and $E_{\rm e}$ is the photon energy at the emission point on the disk, the emission radius of the disk $r_{\rm e}$, and the emission angle in the rest-frame of the material of the disk $\vartheta_{\rm e}$. At the end of these calculations, we have the redshift image of the accretion disk. Since at step~2 we had calculated the non-relativistic reflection spectrum at every radial coordinate, we can infer the spectrum of every pixel of the image of the disk and, integrating over the whole disk image, we find the relativistic reflection spectrum of the source.
\end{enumerate}

These calculations are time-consuming. On the other hand, a model for data analysis has to be able to generate quickly many spectra for different values of the model parameters in order to scan the parameter space and find the best-fit. Current reflection models for data analysis follow a different approach. We write the relativistic reflection spectrum of the source as the following integral (for more details and the approximations behind, see Ref.~\cite{Bambi:2024hhi})
\be\label{eq-Fo}
F_{\rm o} (E_{\rm o}) &=& C \int_{R_{\rm in}}^{R_{\rm out}} dr_{\rm e} \int_0^1 dg^* \, 
\frac{\pi r_{\rm e} g^2}{\sqrt{g^* \left( 1 - g^* \right)}} 
\left[ f^{(1)} ( g^* , r_{\rm e} , i ) + f^{(2)} ( g^* , r_{\rm e} , i ) \right]
\epsilon (r_{\rm e}) \, \bar{I}_{\rm e} (E_{\rm e}) \, ,
\ee
where $C$ is the normalization of the reflection component to be inferred by the fit, $R_{\rm in}$ and $R_{\rm out}$ are, respectively, the radial coordinates of the inner and the outer edges of the disk, $r_{\rm e}$ is the radial coordinate of the emission radius, and $g^* = g^* (r_{\rm e}, i)$ is the relative redshift defined as
\be
g^* = \frac{g - g_{\rm min}}{g_{\rm max} - g_{\rm min}} \, ,
\ee 
where $g_{\rm min} = g_{\rm min} (r_{\rm e} , i)$ and $g_{\rm max} = g_{\rm max} (r_{\rm e} , i)$ are, respectively, the minimum and maximum redshift factors for the emission radius $r_{\rm e}$ and the inclination angle of the disk $i$.

The three key-quantities inside the integral are the transfer functions, $f^{(1)} ( g^* , r_{\rm e} , i )$ and $f^{(2)} ( g^* , r_{\rm e} , i )$, the emissivity profile, $\epsilon (r_{\rm e})$, and the normalized specific intensity of the radiation at the emission point on the disk, $\bar{I}_{\rm e} (E_{\rm e})$:
\begin{enumerate}
\item The transfer function takes all relativistic effects into account and can be calculated at step~3 above~\cite{Cunningham:1975zz}. It depends on the motion of the material in the disk and on the propagation of the photons from the emission point on the disk to the detection point on the plane of the distant observer. The transfer function has two branches, $f^{(1)} ( g^* , r_{\rm e} , i )$ and $f^{(2)} ( g^* , r_{\rm e} , i )$, because for every emission radius $r_{\rm e}$ there are two branches connecting $g^* = 0$ with $g^* = 1$. 
\item The emissivity profile describes the radial profile of the photon flux emitted by the disk and can be calculated at step~1 above, as it is determined by how the corona illuminates the disk 
\item The normalized specific intensity of the radiation at the emission point on the disk can be calculated at step~2 by solving radiative transfer equations. It is ``normalized'' because the actual non-relativistic reflection spectrum at every emission radius should be calculated assuming the actual illuminating spectrum, which, in general, is different at different radii. On the other hand, in Eq.~(\ref{eq-Fo}) we assume the same non-relativistic reflection spectrum over the whole disk (or over a certain part of the disk if we divide the disk into different zones) and we regulate the intensity of the specific intensity of the radiation through the emissivity profile. 
\end{enumerate}

The transfer function, the emissivity profile, and the  normalized specific intensity of the reflection radiation are calculated before the data analysis process and tabulated into FITS files. During the data analysis process, the model simply solves the integral in Eq.~(\ref{eq-Fo}) and calls the FITS files to know the values of the three functions. In this way, it is possible to calculate quickly many reflection spectra by varying the values of the model parameters and we can find the best-fit model for a set of data.


\section{Modeling Simplifications}

Reflection models have been significantly developed in the past 10-15~years, but they still rely on a number of simplifications, so caution is necessary when we analyze very high-quality data and we obtain very precise measurements. The danger is to get very precise -- but not very accurate -- measurements of a source, which can easily lead to an incorrect interpretation of the astrophysical system. It is thus crucial to understand the simplifications in our reflection models and their impact on present and future X-ray observations\footnote{X-ray reflection spectroscopy measurements are inevitably affected even by other systematic uncertainties, like calibration errors, but here we will only consider modeling uncertainties.}.

Modeling simplifications can be roughly grouped into four classes, even if the description of some physical phenomenon can have simplifications from different classes\footnote{For example, the returning radiation is the radiation emitted by the disk and returning to the disk because of the strong light bending near black holes. If we completely ignore the returning radiation, we are in case~4. If we include the returning radiation in the total flux illuminating the disk but we do not take into account that the returning radiation changes the incident spectrum producing the reflection spectrum, we are in case~3; see Subsection~\ref{ss-returning} for more details.}: 
\begin{enumerate}
\item Simplifications in the description of the accretion disk.
\item Simplifications in the description of the corona.
\item Simplifications in the calculations of non-relativistic reflection spectra. 
\item Relativistic effects that are not taken into account or not properly taken into account.
\end{enumerate}

As I will discuss below, in the past years we have extensively investigated the impact of the simplifications in current reflection models on the analysis of present and future high-quality black hole spectra. While several simplifications can sound quite crude and one may expect they can cause large systematic uncertainties, in my opinion current reflection models are not bad and the analysis of current high-quality data can provide precise and accurate measurements of black holes if we select the right sources and the right observations. Most of the modeling uncertainties are under control, even if it is true that there are still some simplifications that, at least for some systems, can be important, their impact on reflection measurements should be understood better, and more advanced models would be useful. In my concern list, there are the following issues:
\begin{enumerate}
\item Emission from the plunging region.
\item Returning radiation.
\item Higher disk electron densities.
\end{enumerate}
The emission from the plunging region and the returning radiation are not taken into account in current model (or included with too crude approximations) and they will be discussed below. The issue of the electron density is that current non-relativistic reflection models may employ electron densities lower than those in real astrophysical systems (especially in the case of stellar-mass black holes in X-ray binaries in soft states). While there have been recent efforts to calculate non-relativistic reflection spectra for higher electron densities, such densities may not be high enough for some real systems. A correlated issue is that current reflection models assume that the electron density is constant along the vertical direction, and it is not clear if this can be a good approximation for the material in the disk. The issue of the electron density will not be discussed below and the reader can refer to \cite{Ding:2024vuy,Liu:2023ovm} and references therein.

The case of future high-quality data is a different story and it is clear that we need more advanced reflection models than those available today to analyze high-quality spectra of bright sources from the next generation of X-ray missions (see, e.g., Subsection~\ref{ss-emiang}).

\subsection{Disk Structure}

Current reflection models normally employ the following assumptions for the structure of the accretion disk:
\begin{enumerate}
\item The disk is infinitesimally thin and perpendicular to the black hole spin axis.
\item The motion of the material in the disk is Keplerian.
\item The inner edge of the disk is at the innermost stable circular orbit (ISCO) or at some larger radius ($R_{\rm in} \ge R_{\rm ISCO}$).
\item There is no emission of radiation from the plunging region. 
\end{enumerate}

In Ref.~\cite{Abdikamalov:2020oci}, we developed a reflection model with a thin disk of finite thickness following the proposal in Ref.~\cite{Taylor:2017jep}. We analyzed a few high-quality, reflection dominated, black hole spectra with the new model and compared the results with the fits obtained imposing an infinitesimally thin disk. Our results suggest that the impact of the thickness of the disk is marginal on the measurements of the model parameters: the measurements with the model with disks of finite thickness are consistent with those obtained with the model with infinitesimally thin disks~\cite{Abdikamalov:2020oci,Tripathi:2021wap,Jiang:2022sqv}.

In Refs.~\cite{Riaz:2019bkv,Riaz:2019kat}, we used ray-tracing techniques to simulate reflection spectra of thick accretion disks. We simulated observations of bright black hole X-ray binaries with \textsl{NICER} and fit the simulated data with an infinitesimally thin disk model for data analysis. While the quality of the fit can be good, some parameters are clearly overestimated or underestimated. Such a result should not be surprising because the relativistic effects from the emission points to the detection points are certainly very different between an infinitesimally thin Keplerian disk and a thick non-Keplerian disk. However, those studies clearly show that caution is necessary in the interpretation of X-ray reflection spectroscopy measurements because infinitesimally thin disks reflection models are commonly used to fit any relativistically blurred reflection spectrum, without worrying about the properties of the accretion disk of the source.

The accuracy of the description of the accretion disk in current reflection models was further investigated in Ref.~\cite{Shashank:2022xyh}. We ran a GRMHD code to simulate thin accretion disks around fast-rotating black holes in the presence of magnetic fields. We used ray-tracing techniques to calculate the reflection spectra of these GRMHD-simulated thin disks and we used the response files of \textsl{NuSTAR} to simulate the observation of a bright black hole X-ray binary. We fit the simulated data with a reflection model employing an infinitesimally thin Keplerian disk and we were able to recover the correct input parameters of the simulations and get acceptable fits. In Ref.~\cite{Shashank:2025} we are extending the analysis reported in Ref.~\cite{Shashank:2022xyh} by running simulations for different values of the black hole spin and its preliminary results are reported in the next subsection.

Current reflection models assume that the disk electron density, $n_{\rm e}$, and the ionization parameter, $\xi$, have the same value at every radial coordinate of the disk. This can sound like a crude approximation, but the idea is that most of the reflection radiation is produced from a relatively small region of the disk, which can be approximated as a region with constant $n_{\rm e}$ and $\xi$. Such an approximation was investigated in Refs.~\cite{Abdikamalov:2021rty,Abdikamalov:2021ues,Mall:2022llu}. In Ref.~\cite{Abdikamalov:2021rty}, we developed a model where the ionization profile is described by a power law and the ionization index can be a free parameter to be fit. In Ref.~\cite{Abdikamalov:2021ues}, we presented a model where the electron density profile is described by a power law, the electron density index can be a free parameter to fit, and the ionization parameter is calculated at every radial coordinate from the emissivity profile and the electron density. We used these two models to fit some high-quality reflection-dominated spectra and we compared our results with those obtained by imposing that $n_{\rm e}$ and $\xi$ are constant. While the fits with the new models with non-trivial electron density and ionization profiles can be better, our analyses do not show significant differences in the estimates of the model parameters, and in particular in the measurements of the black hole spins and inclination angles.

\subsection{Plunging Region}

The plunging region is the space between the inner edge of the disk and the black hole. In the case of a thin Keplerian disk, the accreting material slowly inspirals onto the black hole by losing energy and angular momentum. When it reaches the inner edge of the disk, it quickly plunges onto the black hole. If the density in the plunging region is sufficiently low, the plunging region is optically thin, there is no reflection, and we can see higher order disk images, namely images created by photons that orbited around the black hole one or more times. If the density in the plunging region is not so low, the plunging region is optically thick and we have reflection.

The case of an optically thin plunging region was investigated in Ref.~\cite{Zhou:2019dfw}. It turns out that the reflection radiation from higher order disk images does not appreciably change the relativistic reflection spectrum, especially when the inner edge of the disk is very close to the black hole (and therefore the plunging region is very small), which is the situation is which we can get more precise measurements of the model parameters.

The case of an optically thick plunging region was investigated in Ref.~\cite{Cardenas-Avendano:2020xtw}. If we ignore magnetic fields, the motion of the material in the plunging region should be geodesic~\cite{Bambi:2024hhi}. In such a case, the plunging process is fast, the density of the material is still relatively low, and the material is highly ionized\footnote{The ionization parameter is defined as $\xi = 4 \pi \Phi_{\rm X}/n_{\rm e}$, where $\Phi_{\rm X}$ is the total X-ray flux and $n_{\rm e}$ is the electron density.}. If the material is highly ionized, reflection occurs only through Compton scattering and there is no absorption followed by fluorescent emission. The reflection spectrum from the plunging region is thus a power law spectrum without reflection features, so the analysis of the reflection features in the relativistic reflection spectrum of the source is not appreciably affected by the emission from the plunging region and there is no significant impact on the measurements of key-parameters like the black hole spin or the inclination angle of the disk.

In the presence of magnetic fields, the conclusion is uncertain, even because we do not have currently a good understanding of the magnetic fields around black holes. The role of magnetic fields in the plunging region in current X-ray reflection spectroscopy measurements is under investigation in Ref.~\cite{Shashank:2025} with GRMHD simulations. If the magnetic fields are ``weak'', we recover the results of Ref.~\cite{Cardenas-Avendano:2020xtw}: the material in the plunging region is highly ionized, the reflection spectrum from the plunging region has no reflection features, and in current X-ray reflection spectroscopy measurements we can safely ignore the emission from the plunging region. If the magnetic fields are ``strong'', they slow down the plunging process, the density of the material in the plunging region increases, and its ionization decreases. Now we do have reflection features in the radiation reflected in the plunging region. The result is that a slowly-rotating black hole can mimic a fast-rotating black hole. This effect of the magnetic fields is found in simulations with $\beta_{\rm min} = 0.5$, where $\beta_{\rm min}$ is the minimum value of the radiation pressure to magnetic pressure ratio, which is not normally considered a strong magnetic field and we are not in a magnetically arrested accretion regime (where $\beta_{\rm min} < 0.1$). It is not clear if a model that includes the emission from the plunging region is able to distinguish the spectrum of a fast-rotating black hole from that of a slowly-rotating black hole with reflection features from the plunging region, but this is certainly something that should be investigated soon.

\subsection{Reflection Comptonization}

\begin{figure}[b]
\centering
\includegraphics[width=0.45\linewidth]{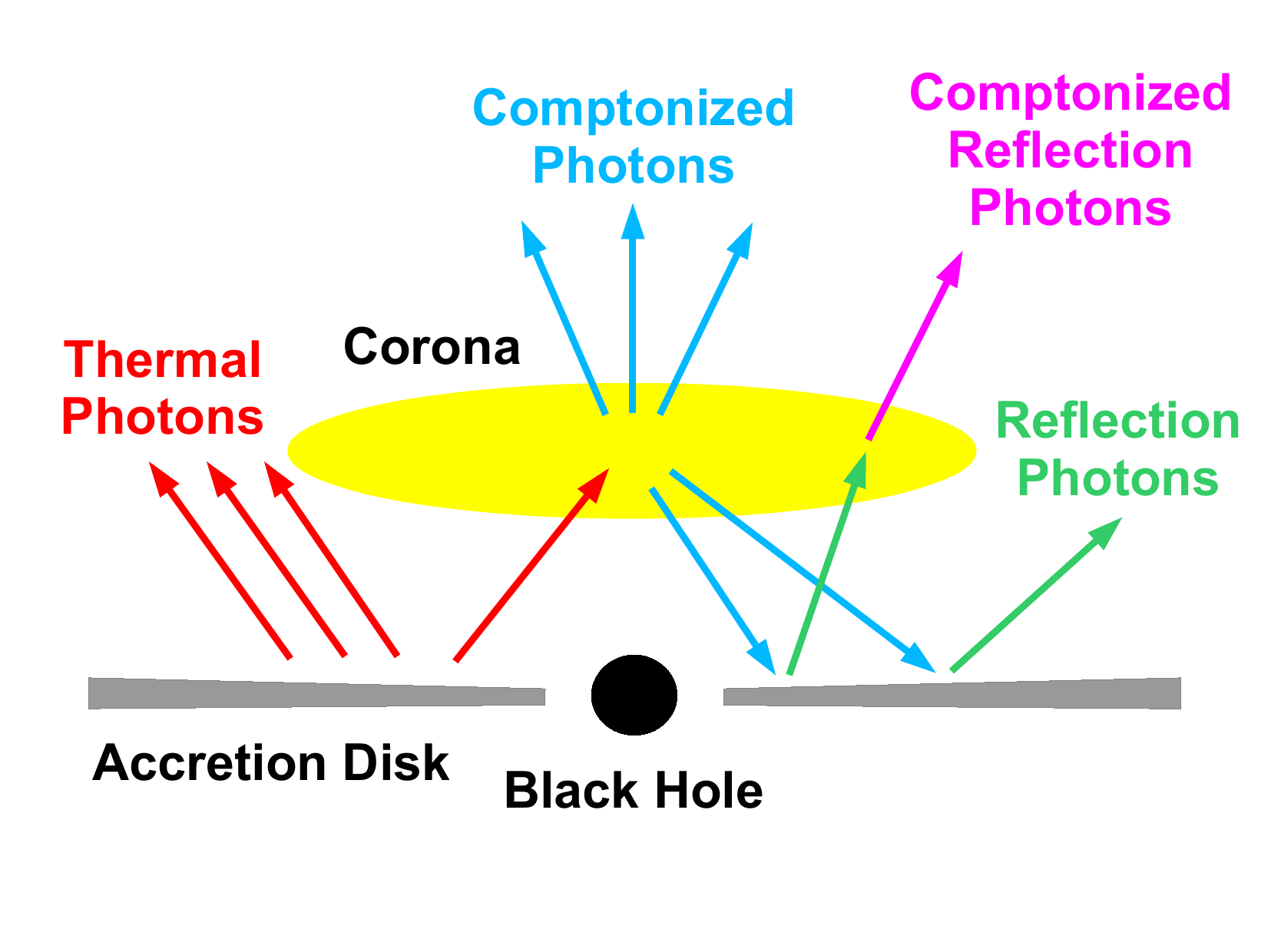}
\vspace{-0.5cm}
\caption{Reflection Comptonization.}
\label{f-refComp}
\end{figure}

If thermal photons from the disk can inverse Compton scatter off free electron in the corona, as illustrated in Fig.~\ref{f-corona}, it is natural to expect that even the reflection photons can inverse Compton scatter off free electrons in the corona. This is the reflection Comptonization and is illustrated in Fig.~\ref{f-refComp}. In Ref.~\cite{Li:2024eue} we analyzed a high-quality \textsl{NuSTAR} spectrum of the black hole X-ray binary EXO~1846--031 to investigate the impact of the reflection Comptonization. While the fit requires that about 10\% of the reflection radiation is scattered by the corona, this does not have any significant impact on the estimates of the main key-parameters of the model, like the measurements of the black hole spin and the inclination angle of the disk.

\subsection{Emission Angle}\label{ss-emiang}

The inclination angle of the disk, $i$, is the angle between the disk axis (which coincides with the black hole spin axis in our models) and the line of sight of the distant observer. The emission angle, $\vartheta_{\rm e}$, is the angle between the normal to the disk and the emission direction of the photon in the rest-frame of the material of the disk. Because of the phenomenon of light bending, $i \neq \vartheta_{\rm e}$ and $\vartheta_{\rm e}$ changes value over the disk. Moreover, the reflection spectrum is not isotropic, so $I_{\rm e} = I_{\rm e} (\vartheta_{\rm e})$.

The non-relativistic reflection model {\tt reflionx}~\cite{Ross:2005dm} cannot calculate reflection spectra for specific emission angles and its output is a non-relativistic reflection spectrum averaged over all angles. Relativistic reflection spectra are obtained by convolving {\tt reflionx} with a convolution model like {\tt relconv}: {\tt relconv$\times$reflionx}. The inclination angle in {\tt relconv} refers to the inclination angle of the disk $i$ and treats the non-relativistic reflection spectrum as if its emission were isotropic.

The non-relativistic reflection model {\tt xillver}~\cite{Garcia:2010iz} can instead calculate reflection spectra for specific emission angles. In the {\tt xillver} FITS file we have thus the parameter $\vartheta_{\rm e}$ (but not the inclination angle of the incident radiation, which is assumed to be $45^\circ$). If we use the model {\tt relconv$\times$xillver}, we can link the value of the inclination angle of the disk $i$ in {\tt relconv} with that of the emission angle $\vartheta_{\rm e}$ in {\tt xillver}, even if the two angles would be different in general.

In the case of the relativistic reflection model {\tt relxill}, we specify the inclination angle of the disk $i$ and the model calculates the averaged emission angle\footnote{In {\tt relxilllp}, which is the {\tt relxill} model for a lamppost corona, we specify the inclination angle of the disk $i$, the model divides the disk into 10~annuli, and it calculates the averaged emission angle for every annulus.}. The model then use the {\tt xillver} spectrum with emission angle equal to the averaged emission angle.

The importance of the actual emission angle in present and future observations is investigated, respectively, in Ref.~\cite{Tripathi:2020cje} and Ref.~\cite{Liu:2025}. In Ref.~\cite{Tripathi:2020cje}, we analyzed very high-quality black hole spectra of current X-ray missions and we fit the reflection components with {\tt relconv$\times$reflionx}, {\tt relconv$\times$xillver}, and {\tt relxill}. We found consistent measurements of the key-parameters of the model, like the black hole spin and the inclination angle of the disk, suggesting that the quality of current X-ray data is not sufficient to distinguish the difference between the inclination angle of the disk and the emission angle.

In Ref.~\cite{Liu:2025}, we are considering future simultaneous observations \textsl{eXTP}+\textsl{Athena} of bright black hole X-ray binaries. With a ray-tracing code, we calculate relativistic reflection spectra by taking the non-relativistic {\tt xillver} spectrum with the correct emission angle at every emission point on the disk. The simulated spectra are then fit with {\tt relxill}. While we can normally recover the correct input parameters, we find large residuals in the best-fit. The current approximation in {\tt relxill} to employ an averaged emission angle does not seem to work well for future high-quality observations and reflection models for the next generation of X-ray mission should certainly improve this part.

\subsection{Returning Radiation}\label{ss-returning}

The returning radiation is the radiation emitted by the disk and returning to the disk because of the strong light bending near black holes. The effect is important only very close to a black hole, so it can be relevant for fast-rotating black holes with untruncated disks, when the corona illuminates mainly the inner part of the disk~\cite{Mirzaev:2024fgd}. The effect of the returning radiation in reflection spectra was investigated with different approximations by different groups in the past~\cite{Dabrowski:1997xr,Niedzwiecki:2007jy,Niedzwiecki:2016ncz,Niedzwiecki:2018gtd,Dauser:2022zwc,Riaz:2020zqb,Riaz:2023xng}. The returning radiation changes the emissivity profile produced by the direct radiation of the corona as well as the spectrum of the radiation illuminating the disk, which deviates from a simple power law with a high-energy cutoff when the returning radiation is taken into account~\cite{Mirzaev:2024fgd,Mirzaev:2024qcu}. The latest version of {\tt relxill} includes the effect of the returning radiation on the emissivity profile, but the non-relativistic spectrum is still calculated assuming that the incident spectrum is a power law with a high-energy cutoff~\cite{Dauser:2022zwc}. As shown in Ref.~\cite{Mirzaev:2024fgd}, for fast-rotating black holes with untruncated disks it is important to calculate the non-relativistic reflection spectra with the correct spectra of the incident radiation. If the non-relativistic reflection spectra are calculated assuming that the incident spectrum is a power law with a high-energy cutoff, the measurements of the model parameters can be affected by large systematic errors. However, it is not straightforward to construct a model for the analysis of black hole X-ray spectra that calculates non-relativistic reflection spectra without the above simplification.

The models presented in Refs.~\cite{Mirzaev:2024fgd,Mirzaev:2024qcu} can calculate non-relativistic reflection spectra with the correct spectra of the incident radiation, because they directly solve the radiative transfer equation in every run, but they are too slow to analyze the spectrum of a source and find a best-fit. As discussed in Section~\ref{s-1}, models for X-ray data analysis calculate the integral in Eq.~(\ref{eq-Fo}) and they get the values of the transfer function, emissivity profile, and normalized specific intensity of the radiation at the emission point from FITS files, where these functions are tabulated. The problem is that the sizes of these FITS files grow very quickly if we increase the number of parameters in the model, and this causes the models to be too slow. In the current version of {\tt relxill}, the FITS file of the normalized specific intensity of the radiation is already 7~GB with six parameters (photon index and electron temperature of the corona to describe the spectrum of the direct radiation from the corona, disk electron density, ionization parameter, and iron abundance to describe the material in the disk, and emission angle). If we wanted to construct a FITS file with the non-relativistic reflection spectra generated by the actual incident spectra, which are the result of the combination of the direct radiation from the corona, the returning radiation of the thermal spectrum of the disk, and the returning radiation of the reflection component itself, we would need a huge FITS file and the process to analyze a spectrum would be too slow.


\section{Towards a New Generation of Reflection Models}

As pointed out at the end of the previous section, it is not clear how we can construct a reflection model for data analysis that takes the returning radiation into account in the proper way. We cannot construct FITS files with many parameters, because this would result in huge files and the data analysis process would become too slow. We cannot calculate the non-relativistic reflection spectra during the data analysis process, as done in the models described in Refs.~\cite{Mirzaev:2024fgd,Mirzaev:2024qcu}, because this would also make the data analysis process too slow. This is not a peculiar problem in X-ray reflection spectroscopy but a very common issue in the analysis of astrophysical data.

Our plan is to try to use machine learning techniques to develop a new generation of significantly more advanced reflection models. As a case study, we want to construct a reflection model that includes the effects of the returning radiation~\cite{Mirzaev:2025}.

The basic idea is to replace current FITS files with neural networks. Instead of using a FITS file to tabulate, for example, non-relativistic reflection spectra, we can train a neural network to predict non-relativistic reflection spectra. After the training process, a neural network is made of files of some MB, even if the model has several parameters. Instead of calling the FITS file, the model for data analysis can call the neural network and quickly extract the non-relativistic reflection spectra necessary to calculate the relativistic reflection spectra to be compared with the observed spectra. The neural network can probably be replaced by a FITS file with the coefficients of the neural network, which can further simplify the model because it would not require applications to work with neural networks.


\section{Concluding Remarks}

X-ray reflection spectroscopy is a powerful tool to study the physics and astrophysics of accreting black holes. In the past 10-15~years, reflection models have been significantly developed and X-ray reflection spectroscopy measurements can be simultaneously precise and accurate if we analyze high-quality data and we properly select the sources and the observations to analyze. In my opinion, there are three effects that need to be understood better and maybe included in current models: emission from the plunging region, returning radiation, and higher disk electron densities. Some effects, like that of the returning radiation, may not be easily implemented in current reflection models. However, we are now trying to use machine learning techniques to develop a new generation of reflection models, with the goal to have models that are at the same time significantly more advanced and still fast to analyze black hole spectra and find the best-fit models.


\acknowledgments

This work was supported by the National Natural Science Foundation of China (NSFC), Grant No.~12250610185, 11973019, and 12261131497, and the Natural Science Foundation of Shanghai, Grant No.~22ZR1403400.


\bigskip
\bigskip
\noindent {\bf DISCUSSION}

\bigskip
\noindent {\bf SILVIA ZANE:} What is the assumption about the seed spectrum producing the Comptonized spectrum of the corona? What is your view on the possibility of extending the code to calculate polarization?

\bigskip
\noindent {\bf COSIMO BAMBI:} Models normally assume that photons with a thermal spectrum (say, temperature $T_{\rm s}$) inverse Compton scatter off free electrons with a thermal distribution (say, temperature $T_{\rm e}$). In such a case, the Comptonized spectrum should be described well by a power law with a high-energy cutoff determined by $T_{\rm e}$ and a low-energy cutoff determined by $T_{\rm s}$. The model {\tt relxill} approximates the Comptonized spectrum with a power law and a high-energy cutoff. The model {\tt relxillCp} employs the Comptonized spectra predicted by the model {\tt nthcomp} for $T_{\rm s} = 1$~eV and different values of $T_{\rm e}$. There are also studies in which a fraction of the free electrons has a non-thermal distribution and one can try to constrain such a fraction of non-thermal electrons with observations.     

Concerning the second question, it is surely possible to extend current models to calculate the polarization of reflection spectra and it is something that people working on reflection modeling want to do in the next years. To do this, $i)$ we need a non-relativistic reflection model (like {\tt xillver}) that includes the calculation of the polarization of the reflection radiation (and as far as I know Javier Garcia is already working to extend {\tt xillver} to include the polarization), and $ii)$ when we solve the geodesic equations to determine the photons trajectories from the emission points on the disk to the detection points on the plane of the distant observer we have also to calculate the rotation of a tetrad along these trajectories (which is straightforward and there are already codes to do these calculations).

\end{document}